\documentclass[apj]{emulateapj}
\usepackage{amsmath}
\usepackage{graphicx}
\usepackage{epsfig}

\usepackage{url}
\usepackage{indentfirst}
\begin{document}
\title{A non-thermal pulsed  X-ray emission of AR~Scorpii}
\author{Takata, J.\altaffilmark{1}, Hu, C.-P.\altaffilmark{2}, Lin  L.C.C.\altaffilmark{3,4},  Tam, P.H.T.\altaffilmark{5}, Pal, P.S.\altaffilmark{5}, Hui, C.Y.\altaffilmark{6}, Kong, A.K.H.\altaffilmark{7} and Cheng, K.S.\altaffilmark{2}}
\email{takata@hust.edu.cn }
\altaffiltext{1}{School of physics, Huazhong University of Science and Technology, Wuhan 430074, China}
\altaffiltext{2}{Department of Physics, The University of Hong Kong, Pokfulam
    Road, Hong Kong}
\altaffiltext{3}{Academia Sinica, Institute of Astronomy and Astrophysics, Taipei, 10617, Taiwan}
\altaffiltext{4}{Department of Physics, UNIST, Ulsan 44919, Korea}
\altaffiltext{5}{School of Physics and Astronomy, Sun Yat-sen University, Zhuhai 519082, China}
\altaffiltext{6}{Department of Astronomy and Space Science, Chungnam National University, Daejeon 34134, Korea}
\altaffiltext{7}{Institute of Astronomy and Department of Physics, National Tsing Hua University, Hsinchu 30013, Taiwan}
\begin{abstract}
 We report the analysis result of UV/X-ray emission  from AR~Scorpii, which is an intermediate
  polar (IP) composed of a magnetic white dwarf and a M-type star, 
with the  \emph{XMM-Newton} data. The X-ray/UV  emission clearly shows a large variation over the orbit, 
 and their  intensity  maximum (or minimum) is located at the superior conjunction (or inferior conjunction) 
of the M-type star orbit.  The hardness ratio of the X-ray emission 
shows a small variation over the orbital phase, and shows no indication of the  absorption  by an accretion column.  
 These properties are naturally explained by the emission  from  the M-type star surface
 rather than from  the accretion column on the WD's star similar to the usual IPs.
 Beside, the observed X-ray emission also modulates with WD's spin with a pulse fraction of $\sim 14\%$.
 The  peak position is aligned in the optical/UV/X-ray band. This supports the  hypothesis that 
 the electrons in AR~Scorpii are accelerated to a relativistic speed, and emit non-thermal photons via
 the synchrotron  
radiation.  In the X-ray bands, the evidence of the power-law spectrum is found in the pulsed component, although  
the observed emission is dominated by the optically thin thermal plasma emissions with several different temperatures.  
It is considered that the magnetic dissipation/reconnection process on the M-type star surface
heats up the plasma  to a temperature of several keV, and also  accelerates the electrons to the relativistic speed.  The  relativistic  electrons are trapped in the WD's closed magnetic field lines by the magnetic mirror effect. In this model,  the observed  pulsed component is explained by the emissions
  from the first magnetic mirror point.    
\end{abstract}
\section{Introduction}

AR~Scorpii (hereafter ARSco) is a white dwarf binary system categorized as the intermediate
polar (hereafter IP), and it is a compact binary system with a
binary separation of  $a\sim 8\times 10^{10}$cm. The distance to the source is  $d\sim 110$pc
(\citealt{ma16,bu17}).  This binary system consists of a magnetic white dwarf, for which the surface magnetic field is $B_s\sim 10^{8}$G,
and a M-type main sequence star (hereafter M-type star) with a radius of $R_*\sim 0.4R_{\odot}$ and a mass of $M_*\sim 0.3M_{\odot}$.
The spin period of the white dwarf is $P_s\sim 117$s, and the orbital period of the system is $P_{o}=3.56$hrs.
In terms of the radio/optical/UV emission properties,
AR~Sco is distinguished  from other  IPs, and is similar to those
of neutron star (NS) pulsar.
The radio/optical/UV emission modulates due to the spin of the WD,  and the light curve
shows the double peak structure. The phase separation between two peaks is
$\sim 0.5$ in the optical/UV bands, and the pulse fraction exceeds 95\%
at the UV bands \citep{ma16}. The optical emission is observed with a strong
linear polarization and a polarization degree varying over the spin phase.  The double peak structure of the pulse profile and the morphology of the linear polarization \citep{bu17} in the optical bands are resemble to those of the Crab pulsar, which is an isolate young NS pulsar with
 electromagnetic wave from radio to high energy TeV bands (e.g. \citealt{ku01, ka05, ta07}). Moreover, the optical emission
from AR~Sco also modulates on the orbital period (3.56hrs), which indicates the heating
of the day-side of the M-type star by  the magnetic
field/radiation of  the WD (\citealt{ma16,ka17}). This feature is also similar to that of the millisecond pulsar/low mass star
binary systems (e.g. \citealt{fr88, ko12}). With the unique properties of the emission, AR~Sco may
be the first WD binary system  that continuously shows a non-thermal
radiation from relativistic electrons.  

AR~Sco's broadband electromagnetic spectrum from radio to UV bands is characterized by a synchrotron radiation from
relativistic electrons, indicating acceleration process in the binary system.
As pointed out by \cite{ge16}, on the other hand, the number of
 particles that emits the observed pulsed optical emission of AR~Sco is significantly larger than the number that can be supplied by the WD itself.
\cite{ge16} thus suggest that an electron/positron beam from
the WD's polar cap sweeps the stellar wind from the M-type star, and a bow shock propagating into stellar
wind accelerates the electrons in the wind. \cite{ta17} consider the relativistic electrons that are trapped
at the closed magnetic field lines of the WD by the magnetic mirror effect, and suggest that the  pulse emission is originated by the emission from the first magnetic mirror point. Both models predict the non-thermal X-ray emission from this system. However, 
\cite{ma16}   report no significance detection of the pulse emission in the X-ray bands, and determine  the upper limit of the
pulse fraction at $\sim 30$\%.  Therefore, the origin of the X-ray emission from this system has not been undetermined.
In this paper, we report the analysis of results from the UV/X-ray data taken by the XMM-Newton. 

\section{Data analysis}
\begin{deluxetable}{ccccc}
  \tablewidth{0pt}
  \tablecaption{Ephemeris of AR~Scorpii
    \label{epheme}}
  \tablewidth{0pt}
  \tablehead{
    $T_{o,ref}\tablenotemark{a}$ & $T_{s,ref}$\tablenotemark{b} & $\nu_b$\tablenotemark{c} & $\nu_s$\tablenotemark{d} &$\nu_o$\tablenotemark{e} \\
    (MJD) & (MJD) & (mHz) & (mHz) & (mHz)}
    579264.09615 & 57641.54629   & 8.4611 & 8.5390 &  0.07792 
  \enddata
  \tablenotetext{a} {Reference time for  the orbital phase. Adopted from \cite{ma16}}
  \tablenotetext{b}{Reference time for the spin phase. }
  \tablenotetext{c}{Beat frequency in Figure~\ref{powerspe}.}
  \tablenotetext{d}{Spin frequency in Figure~\ref{powerspe}.}
  \tablenotetext{e}{Orbital frequency. Adopted from \cite{ma16}} 
\end{deluxetable}

We analyze the archive XMM-Newton data taken at 2016 September 19 (Obs. ID: 0783940101, PI: Steeghs).
This new observation was performed with a total exposures of $\sim$~39ks.
The observation was operated under the fast mode for the  OM camera, 
the small window  mode for the MOS1/2 CCDs (time resolution 0.3s),
and the large window mode for the PN camera (time resolution 47.7ms).
Event lists from the  data are produced  in the standard way using the
most updated instrumental calibration, $omfchain$, $emproc$ and $epproc$ tasks
of the  XMM-Newton Science Analysis Software (XMMSAS, version 16.0.0).
A  point source is significantly detected ($>100\sigma$) by the XMMSAS task $edetect\_chain$
at the position of AR~Sco. To perform the spectral and timing analyses,
we extract the EPIC data from a circle region with a radius of $20''$  centered
at source position $({\rm R.A.}, {\rm Decl.})=(16^{\rm h}21^{\rm m}47^{\rm s}.29, -22^{\circ}53'10''.4)$ (J2000). The arrival times of all the
selected events of the OM/EPIC data are barycentric-corrected with the aforementioned
position and the latest DE405 Earth ephemeris.

\subsection{Timing analysis}
\label{timing}
\begin{figure}
  \centering 
  \epsscale{1}
  \includegraphics[scale=0.5]{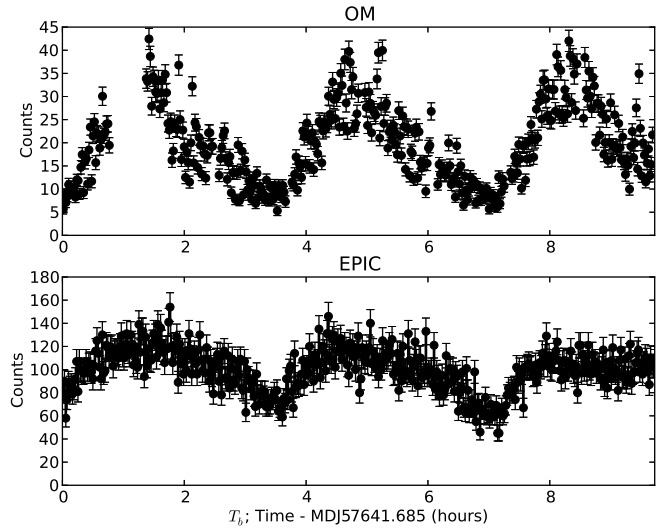}
  \caption{Light curves with a timing resolution of 80s for the OM (top) and all the EPIC data (bottom) after
    background subtraction.  }
  \label{light-xuv}
\end{figure}

\begin{figure}
  \centering
  \epsscale{1}
  \includegraphics[scale=0.5]{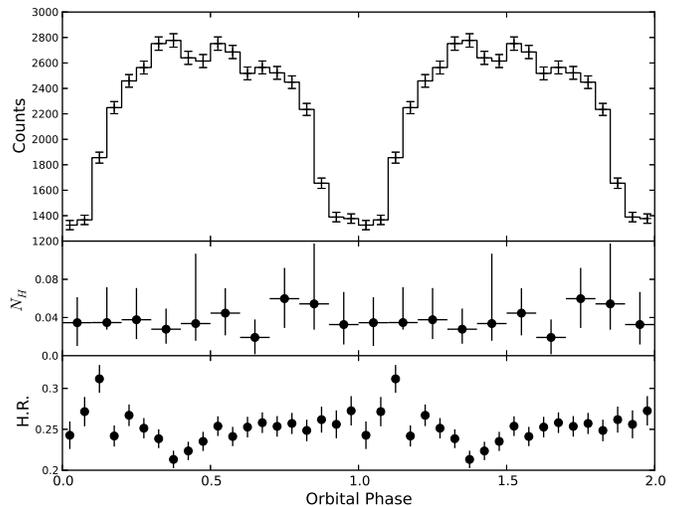}
  \caption{Folded light curve of the EPIC data with $P_b=3.56$hours (top)
    and the evolution of $N_H (10^{22}{\rm cm^{-2}}$ (middle) and
    hardness ratio (bottom), which is defined
    by H.R=$N_{2-12{\rm keV}}/N_{0.15-2{\rm keV}}$.  }
  \label{light-ob1}
\end{figure}

\begin{figure}
  \centering
  \epsscale{1.0}
  \includegraphics[scale=0.5]{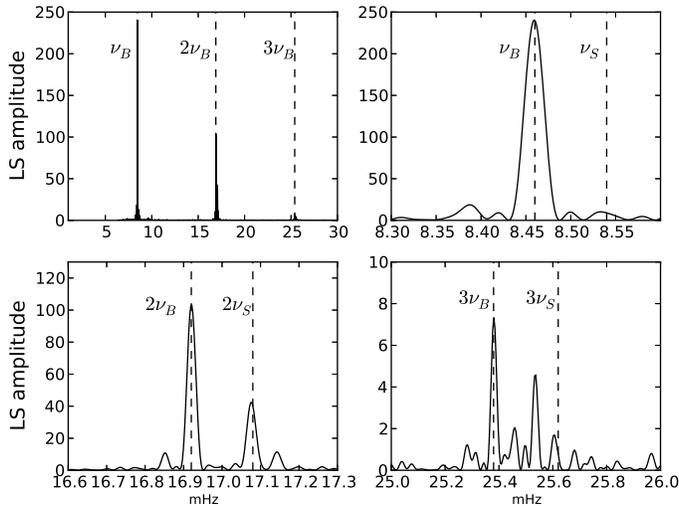}
  \caption{Lomb-Scargle periodogram of the OM data. The location of beat frequency ($\nu_B\sim 8.46$mHz), 
    the spin frequency ($\nu_s\sim 8.54$mHz)
    and their harmonics are indicated by the vertical dashed lines. }
  \label{powerspe1}
\end{figure}

\begin{figure}
  \centering
  \epsscale{1.0}
 \includegraphics[scale=0.5]{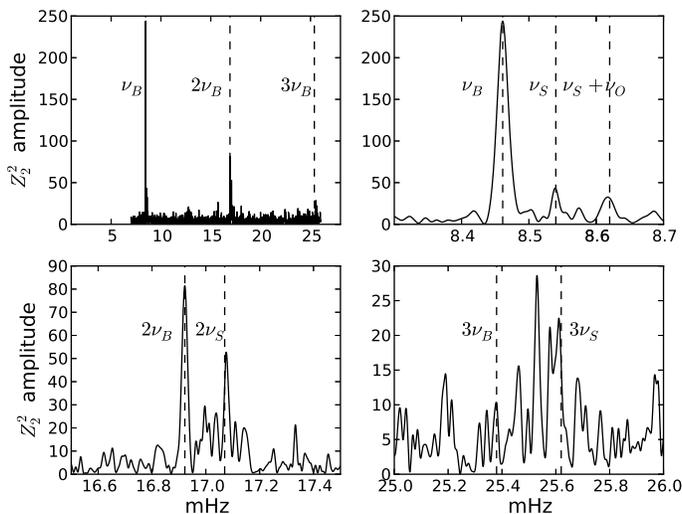} 
 \caption{$Z_2^2$  periodogram for the EPIC data.}
  \label{powerspe}
\end{figure}

\subsubsection{Orbital modulation}

\cite{ma16} argue that   the
un-pulsed optical/UV emission shows  the maximum (or minimum) brightness 
at  the superior conjunction (or inferior conjunction) of the M-type star orbit, and the emission
is originated from the day-side of the M-type star.
The new  \emph{XMM-Newton} observation covers more than two orbits of AR~Sco, and the X-ray
emission significantly modulates over the orbital phase (Figure~\ref{light-xuv}).
We can see in the figure that the X-ray modulation after subtracting the background
remains a large  DC level, and it  synchronizes with the UV orbital modulation. 

We fold the EPIC data in the orbital frequency $\nu_{o}=0.07792$mHz (Table~1) with the reference time $T_o$ (MJD)=579264.09615, where
the M-type star is in  the inferior conjunction; that is, the  M-type star is located between the WD and Earth. The property of the X-ray
orbital modulation of AR~Sco is distinguished
from those of other  IPs. AE~Aquarii is the  IP system whose  orbital period ($P_o\sim 9.88$hours)
and spin period ($P_s\sim 33$s) are similar to those of AR~Sco (\citealt{ch99,it06, te08}),
but its X-ray emission does not show significant orbital modulation.
The observed X-ray flux from some IPs shows a
sharp drop to zero due to  an eclipse of the emission region, which indicates that the X-ray emission region is confined close to the WD's surface \citep{cr02}.  Some  IPs exhibit an orbital modulation due to an 
absorption by an  accretion stream, for which the $N_H$ and
hardness ratio in X-ray
bands rapidly vary  with the change of the observed
flux (\citealt{eva04,pe11,re17}).  For AR~Sco, however, a  large DC component of  the observed light curve
suggests that the size of the  X-ray emission region is comparable to the size of the
binary system. Moreover,  the  small variation of the hardness of the X-ray emission
over the orbit (Figure~\ref{light-ob1}) indicates that the absorption by the accretion matter 
is not the origin of the observed orbital modulation.  Instead, the large variation of the orbital modulation with
a DC component and synchronizing with the UV modulation is  naturally explained
if  the emission is originated from the day-side of the M-type star,  on which the plasma
is  heated by  the magnetic field/radiation of the  WD. 

In Figure~\ref{light-xuv}, we find that OM orbital light curve shows a faster rise and a slower decay, and the orbit maximum is prior
to the superior conjunction of the companion orbit. This orbital waveform is consistent with the previous results (\citealt{ma16,li17}).
This orbital shift is interpreted as a consequence of either (1) the major magnetic dissipation at the leading surface of the M-type
star or (2) the precession of the rotation axis of the WD owing to a misalignment to the orbital axis \citep{ka17}. \cite{li17} reveals
that the orbital waveform and maxim gradually shift with time. 

\subsubsection{Energy dependent pulse profile}
\begin{figure}
  \centering 
    \epsscale{1.0}
    \includegraphics[width=0.5\textwidth]{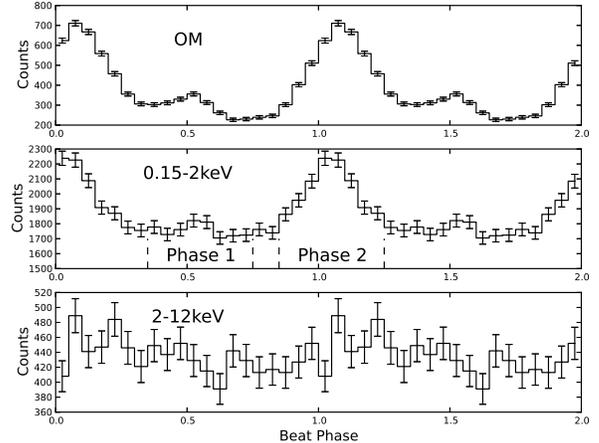}
    \caption{Energy dependent pulse profiles folded in beat frequency: UV (top panel),
      0.15-2keV energy bands (middle panel) and 2-12keV energy bands (right panel). For the  OM data,
      we remove the data at $T_b<3$hours in Figure~\ref{light-xuv},
      since there is a large  observational gap. The spectrum of the pulsed component is generated by subtracting the spectrum at 'Phase 1' from that
    at 'Phase 2'}
  \label{light}
\end{figure}

In \cite{ma16}, the timing analysis shows that
the optical/UV emission from AR~Sco is modulating with the beat frequency of the WD's spin frequency  and the orbital
frequency. The pulse profile folded in the beat frequency  shows a
double peak structure with a  phase-separation of $\sim 0.5$. Moreover,
the peaks in the optical and UV bands are in phase, and the radiation power of the pulse emission in the optical/UV bands is comparable to or more than the DC  emission from the M-type star surface and WD surface. The broadband spectrum from the radio to optical bands
is described by a non-thermal spectrum. These optical/UV properties suggest that the  AR~Sco
continuously generates relativistic electrons.
To confirm  this hypothesis, the detection of the pulse emission in a higher energy band and
the correlation of the pulse peaks in different energy bands are  important. 
We therefore search the beat frequency ($\nu_{B}\sim 8.6$mHz) reported by \cite{ma16} in the  OM/EPIC data, 
and we find a significant peak at the beat frequency in the Lomb-Scarge
periodogram \citep{lo76} in  OM data (Figure~\ref{powerspe1}),
and in the $Z^2_2$ periodogram  \citep{bu83}  in all EPIC data (Figure~\ref{powerspe}).   
For the EPIC data, the beat frequency [$\nu_{b}=8.461100(8)$mHz, where the error is determined by
  the equation (6a) in Leahy (1987)] is detected with  $Z_2^2~\sim 240$ or $H\sim 242$ of the H-test \citep{de10}, which corresponds to
 random probability of $<10^{-14}$, 
suggesting the X-ray pulsation is significantly detected.
In addition to the fundamental beat frequency, 
a peak in periodogram can be found at  the spin frequency [$\nu_s\sim 8.5390$mHz],
and their  harmonics. Table~1 summarizes the ephemeris used to make
a folded light curve in this paper. 

To investigate property of  the pulse profile, we fold the OM/EPIC data into 
the beat phase, and obtain the orbitally phase averaged pulse profiles in the UV bands, 
0.15-2.0keV bands, and 2-12keV  bands (Figure~\ref{light}).  
 In the  UV bands,  the pulse profile is composed 
of the  prominent  first peak and a small second peak, and the phase-separation 
between the peaks is $\sim 0.5$ in the beat phase, which is consistent with the previous 
result of the optical/UV pulse profiles \citep{ma16}. 
We can see in Figure~\ref{light}, the pulse profile  in 0.15-2keV bands is similar to that of  UV bands, 
although the second peak is less significance ($<3\sigma$).  This similarity in the pulse profile shows that the pulse emission in the
UV/soft X-ray bands  is produced by the same population of the particles.
 
The narrow phase width of the main peak and the double peak pulse profile  of
AR~Sco are
also distinguished  from the pulse profiles 
 of canonical IPs, in which the X-ray spin modulation
 is observed as a broad single pulse (e.g. \citealt{pe11}).
 This observational fact also supports the hypothesis that 
 the X-ray emission of AR~Sco is not explained by the emission from the
 accretion column/heated  WD's surface. Our result shows  that the pulse emission from AR~Sco extends
 from radio to soft X-ray bands, and the pulse profile  is aligned  
 from the optical/UV to X-ray energy bands (three to four order of magnitude in the  energy). This is
 a strong indication  that the electrons are accelerated to relativistic energies   in AR~Sco system, and the pulse emission is produced by the synchrotron radiation process.  
In 2-12keV energy band, the detection of the pulsation is not significant with $\chi^2/{\rm dof}=50/24$ for the probability of a flat distribution.

Another interesting feature is the energy dependent pulse fraction.
The pulse fraction, which is defined by the equation  $(f_{max}-f_{min})/(f_{max}+f_{min})$,
 is  measured as $\sim 50$\%  in the UV bands, while it is  $\sim 14\%$ in  0.15-2keV energy bands, 
 which is consistent with the upper limit of $30\%$ in \cite{ma16}. The small pulse fraction shows 
 that the DC level is the main component in the X-ray emission. Moreover, the large orbital variation synchronizing with the
 UV emission shows that  DC component is originated from the heated-side of the M-type star, 
and there is a $\sim$ keV  plasma  around the M-type star surface.
The energy conversion from
the magnetic energy to the particle  energy due to  
the  magnetic reconnection/dissipation on the M-type star is a possible scenario  to heat up/accelerate the plasma.

\subsubsection{Pulse profile; orbital evolution}
\begin{figure*}
  \begin{center}
    \begin{minipage}{0.49\linewidth}
      \includegraphics[width=0.95\textwidth]{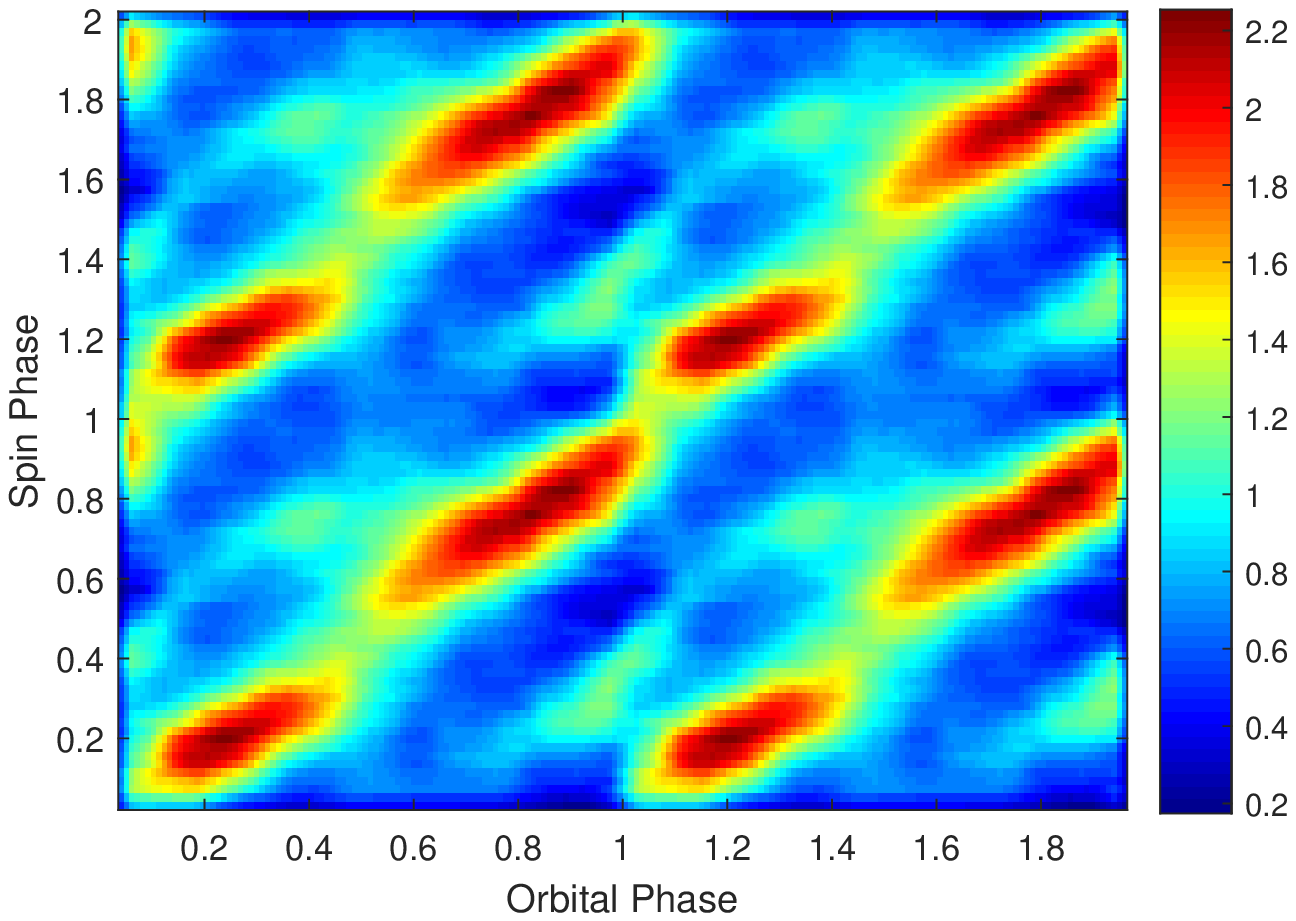}
      \includegraphics[width=0.95\textwidth]{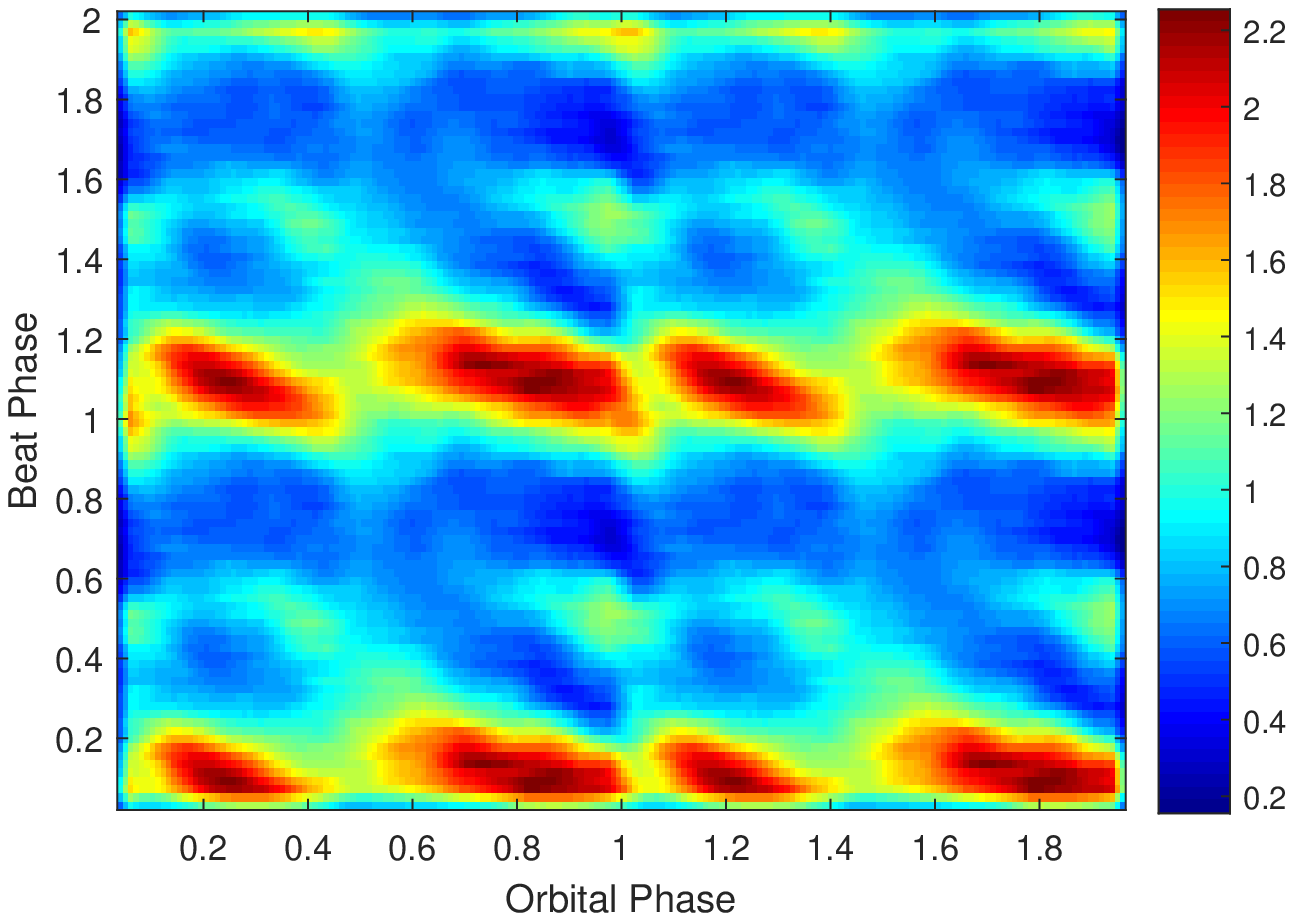}
    \end{minipage}
    \begin{minipage}{0.49\linewidth}
      \includegraphics[width=0.95\textwidth]{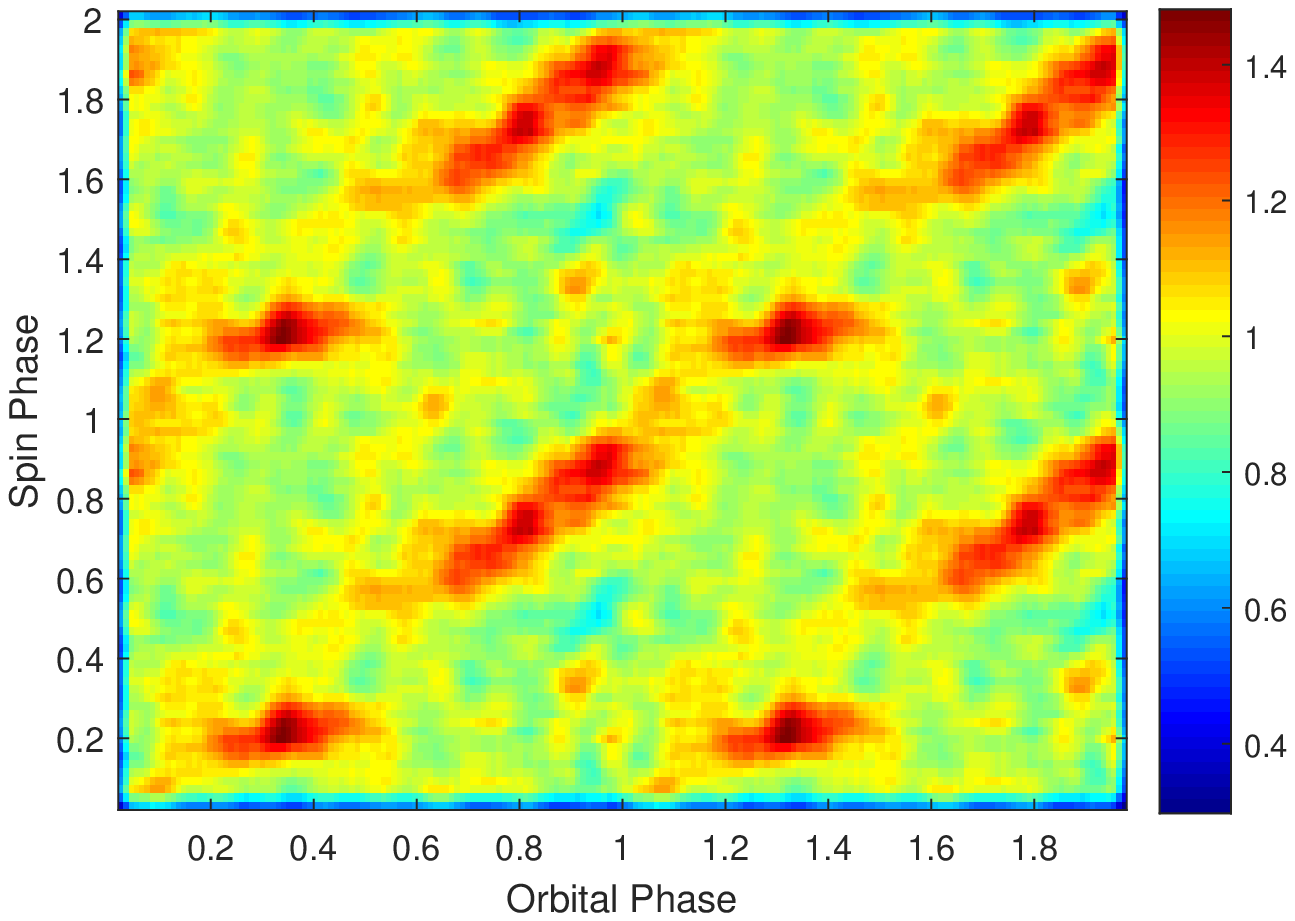}
      \includegraphics[width=0.95\textwidth]{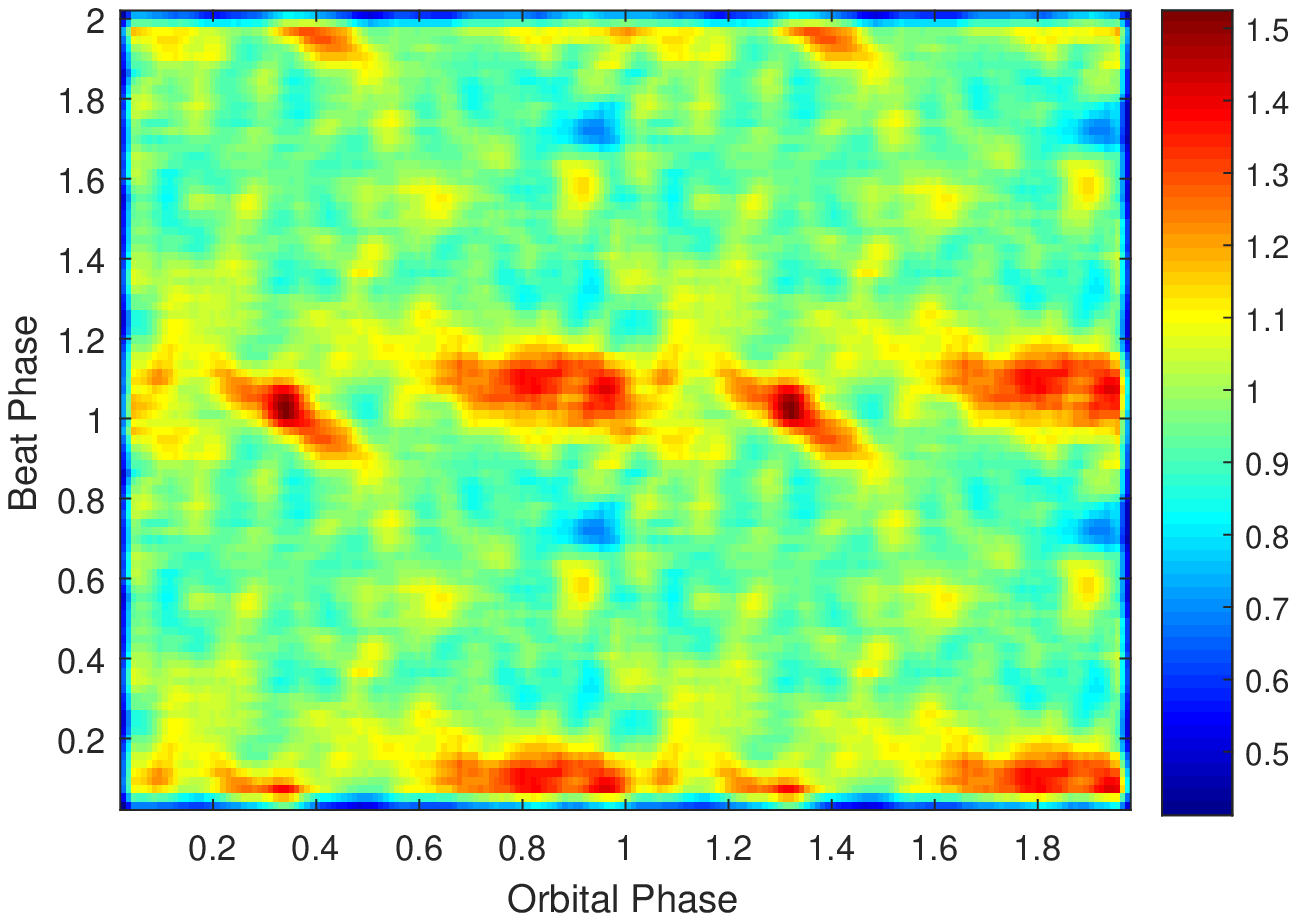}
    \end{minipage}
  \end{center}
  \caption{Dynamics pulse profiles for the OM (left) and for EPIC (right) data.  The data are  folded in
  spin phase (upper panel) and in beat phase (lower panel). }
\label{dynamics}
\end{figure*}

\begin{figure*}
  \begin{center}
    \begin{minipage}{0.49\linewidth}
      \includegraphics[width=1.\textwidth]{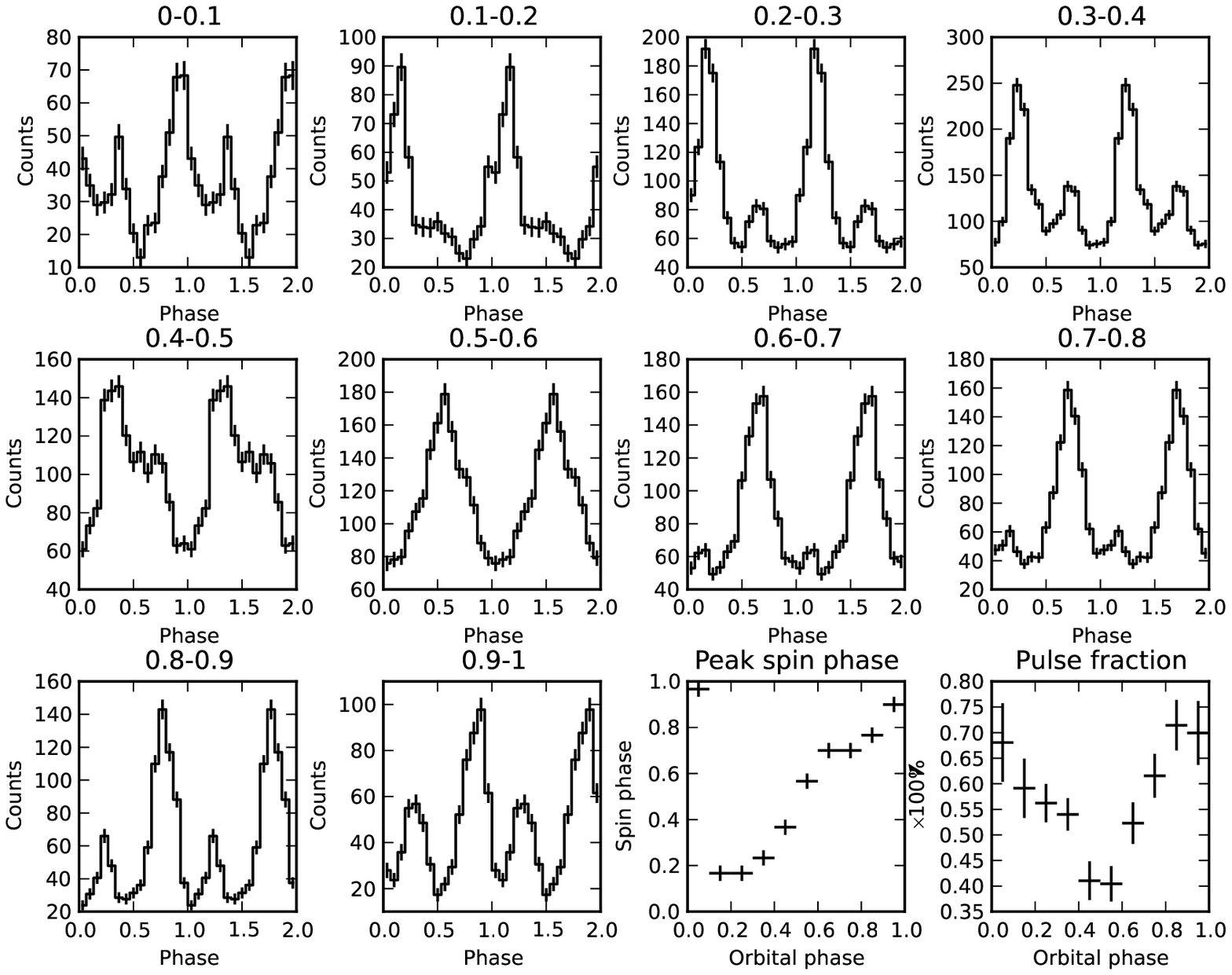}
    \end{minipage}
    \begin{minipage}{0.49\linewidth}
      \includegraphics[width=1\textwidth]{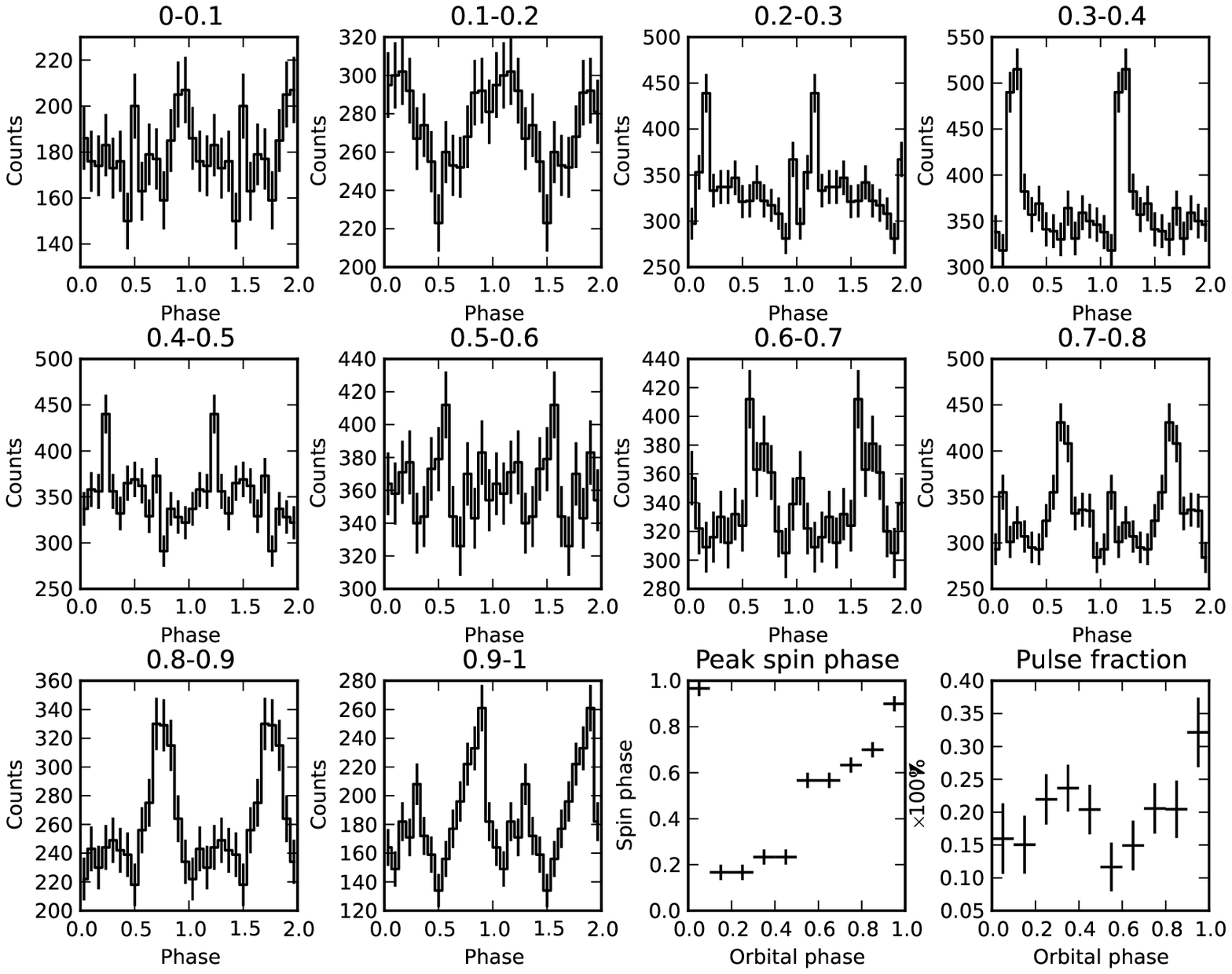}
    \end{minipage}
  \end{center}
  \caption{Orbital resolved pulse profiles for the OM (left) and  EPIC (right) data, after subtracting 
  the   background.  The data are folded in the spin phase. For OM data, we remove the data at $T_b<3$hours in Figure~\ref{light-xuv},
    since there is a large observational gap.}
  \label{pulse-ob}
\end{figure*}

The pulse emission from AR~Sco is observed with the beat frequency $\nu_B$. This indicates that
 the position of the pulse peak in the spin phase has  a  linear shift in  the orbital phase. 
To confirm this, we make a dynamic pulse profile folded in the  spin  phase over the orbital phase (upper panel in
Figure~\ref{dynamics}).  In Figure~\ref{dynamics}, we can clearly see the  
shift of the position of the pulse peak in the spin phase for both OM (right panel) and
EPIC (left panel) data. However, an interesting feature  can be seen in the dynamic pulse profile of 
the EPIC data; in the right upper panel, the phase shift of the pulse peak at
$\Phi_{ob}\sim 0.2-0.5$ orbital phase
is slower than that in $\Phi_{ob}\sim 0.5-1$ orbital phase, and results in a discontinuity of the peak position at the superior conjunction ($\Phi_{ob}\sim 0.5$)
and inferior conjunction ($\Phi_{ob}\sim 0$) of the orbit of the M-type star.
This interesting feature can also be seen in the dynamic 
pulse profiles folded in the beat phase (lower panel in Figure~\ref{dynamics}); in this case,  
the X-ray pulse peak (right pane)  shifts at the $\Phi_{ob}\sim 0.2-0.5$ orbital phase, while the peak position does not 
show a  shift during the  $\Phi_{ob}\sim 0.5-1$ orbital phase. 

To investigate an evolution of the pulse profile over the orbit,
we create orbitally resolved  pulse profiles, folded in spin phase, 
of the OM data and the EPIC data (Figure~\ref{pulse-ob}). In Figure~\ref{pulse-ob}, we can see that the pulse shapes  in
both UV and X-ray bands evolve over  the orbital phase.  On the other hand,
we can also see that the pulse shape and peak position in the UV/X-ray bands
are similar each other in most of the orbital phase.  This observational result
also supports the hypothesis that the pulse emissions in  the UV
and X-ray bands are originated from the same population of the particles.

In the UV bands, the double peak structure can be clearly seen at most part of the orbital phase.
Around $\Phi_{ob}\sim 0.5$ orbital phase (superior conjunction), however, the pulse profile
is described by a broad single peak, and the pulse shape drastically changes during $\Phi_{ob}=0.3-0.6$ orbital
phase, where the shift of the main peak position in the spin phase
is faster than other orbital phase. In the X-ray bands, we also confirm such a rapid change in
the behavior of the pulse profile around the superior conjunction. 
In Figure~\ref{pulse-ob}, a large change in the pulse profile  can be also seen
around the inferior conjunction ($\Phi_{ob}\sim 0$).

\begin{figure}
\centering
\includegraphics[scale=0.4]{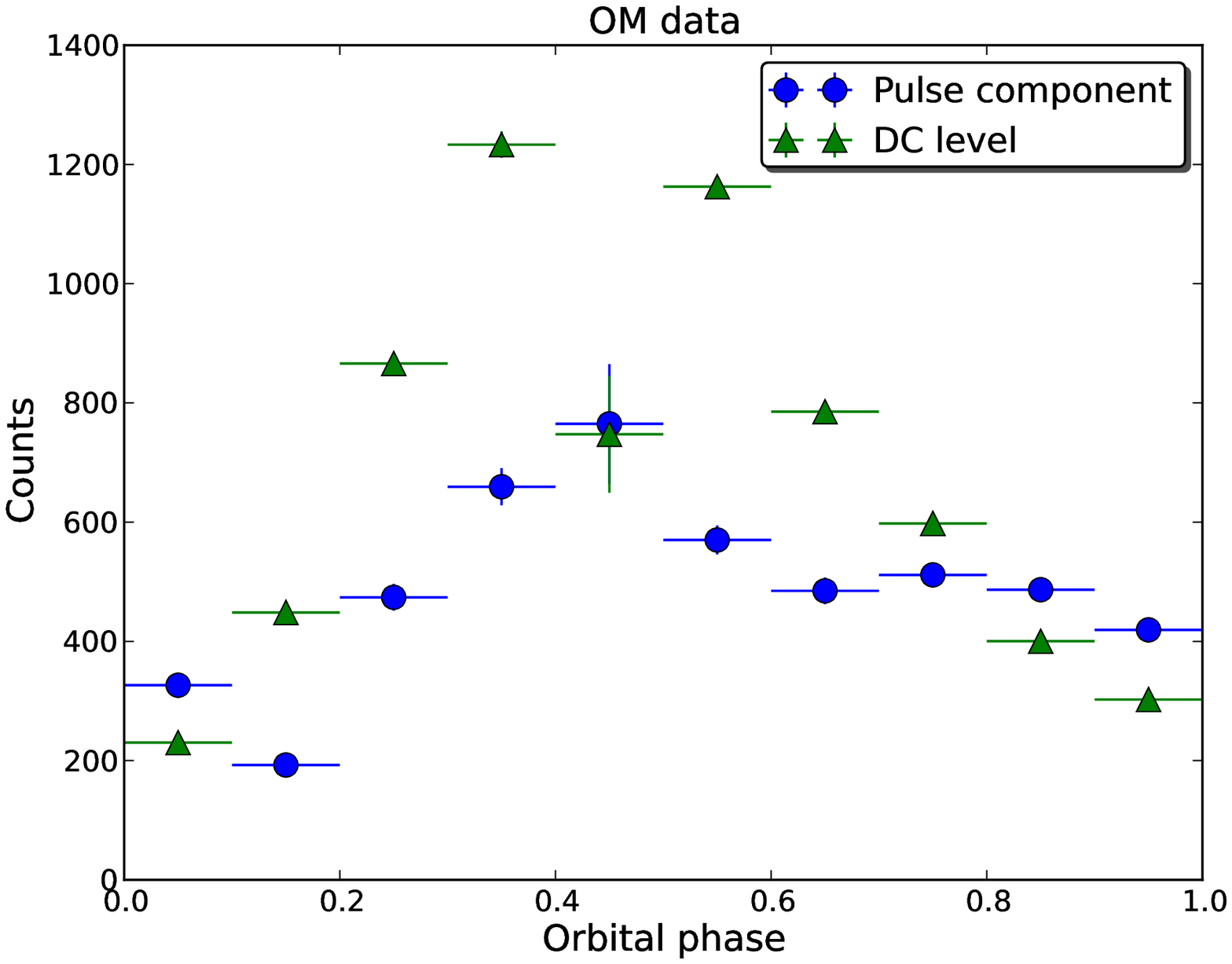}
\caption{Evolution of the pulsed component (circle) and DC level (triangle)  over  the orbital phase for
the OM data.}
  \label{puldc}
\end{figure}

An interesting  feature is the large variation of the pulse fraction over the orbital phase. For
the UV bands, the pulse fraction is maximum at the inferior conjunction
($>70$\%) and minimum  at the superior conjunction ($\sim 40$\%).
The   X-ray emission also  shows a similar trend,
although the uncertainty is large.
For the OM data, we fit the pulse profile with two Gaussian components, and determine the DC level
for each orbital phase. In Figure~\ref{puldc}, we can see that
the DC component varies by  a factor of 6 over  the
orbital motion, while the pulsed component  changes by a factor of $\sim 4$.
For the pulsed component, moreover, the count is almost constant during $\Phi_{ob}=0.4-1.0$ orbital phase.
The difference in the evolution of the photon count over the orbital phase suggests that
the emission region of the pulsed component and DC level is  different.
Since the DC level emission is likely came from the entire surface of the day-side of the M-type star,
the pulsed component is produced in  part of the M-type star's surface or at the WD's magnetosphere.

\subsection{Spectral analysis}
\label{sectrum}

\begin{deluxetable}{lcc}
\tablewidth{0pt}
\tablecaption{Best-fit parameters of the two and three-temperature model.
  \label{fitmodel}}
\tablewidth{0pt}
\tablehead{
   & \colhead{2VMEKAL} & \colhead{3VMEKAL}
}
\startdata
$N_{\rm H}$ ($10^{20}{\rm cm^{-2}}$) & $3.5^{+0.5}_{-0.5}$ & $3.4^{+0.8}_{-0.8}$ \\
$kT_1$ (keV) & $8.0^{+1.8}_{-1.4}$ & $8.0^{+2.8}_{-1.6}$ \\
$kT_2$ (keV) & $1.1^{+0.14}_{-0.17}$ & $1.7^{+0.42}_{-0.26}$ \\
$kT_3$ (keV) & - & $0.6^{+0.08}_{-0.09}$ \\
$N_1$\tablenotemark{a} $(10^{-4})$ & $7.6^{+3.0}_{-3.5}$ & $6.1^{+2.6}_{-2.1}$  \\
$N_2$\tablenotemark{a} $(10^{-4})$ & $0.83^{+0.45}_{-0.35}$ & $2.1^{+1.4}_{-0.9}$ \\
$N_3$\tablenotemark{a} $(10^{-4})$ & -  & $0.35^{+0.15}_{-0.13}$ \\
$F_e$\tablenotemark{b} & $0.62^{+0.48}_{-0.21}$ & $0.67^{+0.29}_{-0.17}$ \\
$F_{X}$\tablenotemark{c} ($10^{-12}~{\rm erg~s^{-1}cm^{-2}}$) & $3.2^{+0.07}_{-0.07}$ & $3.2^{+0.1}_{-0.1}$ \\
$\chi_{\nu}^2$ (dof) & 459 (407) & 416 (404 ) 
\enddata
\tablenotetext{a} {Normalization of the VMEKAL component in units of $10^{-14}/(4\pi d^2)\int n_en_HdV$, where $D$ (cm) is the distancde to the source.}
\tablenotetext{b}{Solar abundances by Anders and Grevesse (1989).}
\tablenotetext{c}{Unabsorbed flux in 0.15--12~keV.}
\end{deluxetable}

\subsubsection{Phase averaged spectrum}
\begin{figure}
  \centering
  \epsscale{0.8}
  \rotatebox{-90}{\includegraphics[scale=0.3]{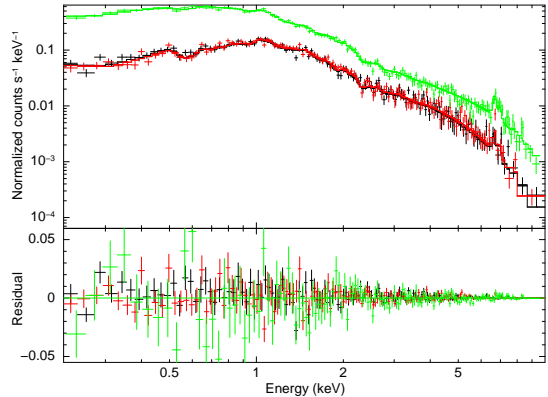}}
  \rotatebox{-90}{\includegraphics[scale=0.3]{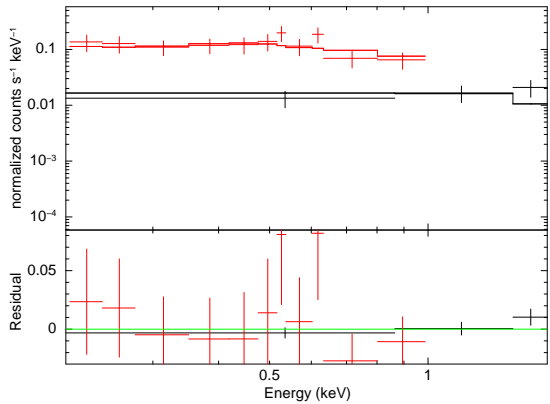}}
  \caption{ X-ray spectra of AR~Sco. Top; phase-average spectrum. Bottom; spectrum of the pulsed component. }
  \label{spectrum}
\end{figure}

In order to further investigate the X-ray emission from AR~Sco,
we carry out a spectral analysis with the  EPIC camera.  We generate the spectra from  photons
in the 0.15-12keV energy bands within a radius of 20'' circle centered at the source. The background spectrum is  generated from a source free region.
The response files are generated by the XMMSAS tasks $rmfgen$ and $arfgen$. We group the channels so as to archive the signal-to-noise ratio S/N $\ge$ 3
in each energy bin with $specgrop$ of SAS, and use Xspec (version 12.9.1) to fit the data.

The obtained spectra (Figure~\ref{spectrum}) clearly  show a  $6.8$keV  emission line from  He-like $F_e$. 
To fit the EPIC data, therefore,  we  adopt an  optically thin thermal plasma emission (VMEKAL in Xspec), which is 
a common spectral model for the IPs.  During the fitting, we find that the current EPIC data can constrain only the abundance of $F_e$, and therefore 
we fix other elements at the solar abundance.   First, we fit the data with a single temperature model, and find that  the model cannot provide an acceptable fit ($\chi^2\sim 713$ for 410 dof).  Adding a power-law component or a  disk component ($diskbb$ in Xspec) does not improve the results of the fitting.  Then we fit the data with  two different temperature components, and find that a two component model
with $kT_1\sim 8.0$keV  and $kT_2\sim 1.1$keV can provide an acceptable fit (Table 1). In order to
determine the number of VMEKAL components with different temperatures, we fit the data with three temperature  and four temperature  component models. An improvement of the fitting is found by adding the  third component with a $F$ statistic  value of 14.1, which means that the probability of this improvement being caused by chance is $1.2\times 10^{-8}$.   Less significant improvement
is found for a fourth component with a  $F$-value of 1.8 or a change probability of 0.15.  The best-fit two and three  temperature VMEKAL models are shown 
in Table~1. We do not find significant evidence of a non-thermal component 
in the phase-averaged spectrum.
Since the  X-ray flux modulates over   the orbit (Figure~\ref{light-xuv}),  
we  evenly divide an  one orbit into four parts to investigate the evolution of the spectrum over  the orbital phase.
  We do not find any significant change in the orbitally phase-resolved spectra.

\subsubsection{Spectrum of the pulsed component}
It has been considered that the emission from  radio to UV bands is produced 
by the synchrotron radiation of the relativistic 
electrons \citep{ma16}.  As described in section~\ref{timing}, the pulse UV/X-ray 
emission  is originated from the same population of the plasma.  
To examine the contribution of the non-thermal component in the 
X-ray bands,   we first perform a phase-resolved spectral analysis, and
we compare  the spectra at  'Phase1'  and at 'Phase 2' in Figure~\ref{light}. We find that the phase-resolved spectra
 for Phase 1/Phase 2  are well described by an  optically thin thermal 
plasma emission model,  and we do not find 
significant difference in the fitting parameters of the two phases 
within $1\sigma$ error. This would be because the pulse fraction is 
$\sim 14$\% and therefore the phase-resolved  spectrum is also
dominated by the  un-pulsed component.

\begin{figure}
    \epsscale{1.0}
    \includegraphics[width=0.5\textwidth]{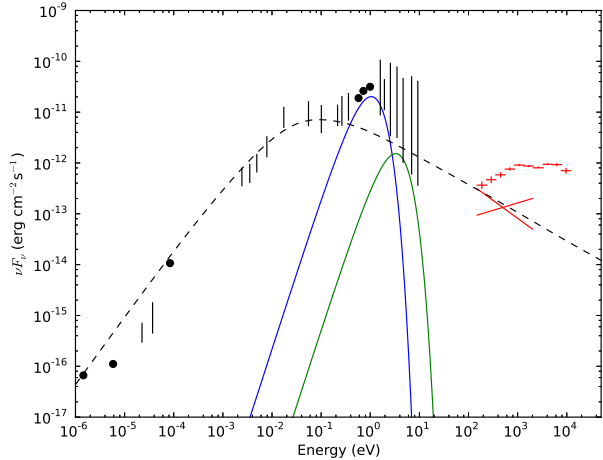}
    \caption{Spectral energy distribution of AR~Sco. The red crosses and lines represent the spectrum of the time average and of the pulsed
      component respectively. The radio/IR/Optical/UV data were taken from \cite{ma16}. The blue and green lines show
      the blackbody spectra for the M-type star ($R_*=0.36R_{\odot}$, $T_*=3100$K) and for the WD ($R_{WD}=0.01R_{\odot}$, $T_{WD}=97500$K),
      where we ignore the absorption of the blackbody emission by the stellar
      atmosphere. The dashed line show the model
      spectrum of the synchrotron emission by assuming the power law index of the injected electrons of $p=3$  [see \cite{ta17} for a detail].}
  \label{sed}
\end{figure}

 We then generate  the spectrum of the pulsed component by subtracting 
 the spectrum of 'Phase 1'  from that of 'Phase 2'. We remove the MOS1 data because of a small amount of the photon counts.
  During  the fitting, we fix the hydrogen column density at $N_H=3.5\times 10^{20}
 {\rm cm^{-3}}$ from the phase-averaged spectrum.
As a result, the power-law model can describe the data reasonably
well ($\chi^2=8.97$ for 12 dof). It yields a  soft emission with a
photon index $\Gamma=2.3\pm 0.5$ and an unabsorbed
flux of $F_{X}=3.7^{+0.7}_{-0.6}\times 10^{-13}{\rm erg~cm^{-2}s^{-1}}$ in 0.15-2keV energy bands.
An optical thin plasma emission model (1 MEKAL model) also results in a comparable goodness
of the fit. In this model, however, the fitting cannot constrain the parameters
of the model, that is, the error range is larger than the central value. Figure~\ref{sed} shows the broad-band SED 
spectrum of AR~Sco. 

\section{Discussion}
\begin{figure}
\epsscale{1.0}
\includegraphics[width=0.5\textwidth]{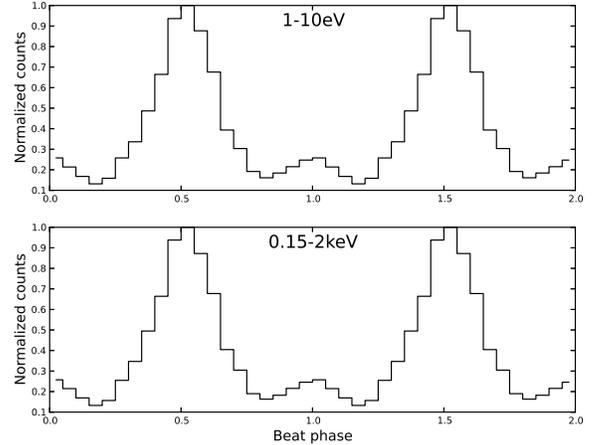}
\caption{Model pulse profiles folded in the beat frequency. Top: 0.1-1 eV energy bands. Bottom: 0.15-2keV energy bands.}
  \label{pulmodel}
\end{figure}

With $d=110$pc, the X-ray luminosity is of the order of $L_{X}\sim 4\times 10^{30}{\rm erg~s^{-1}}$, similar to $L_{X}\sim 10^{31} {\rm erg~s^{-1}}$ of AE~Aquarii \citep{ki14}, but it is two to three orders of magnitude lower than that of typical IPs.  One interesting
feature of AR~Sco is the weak absorption of the
soft X-rays, which corresponds to $N_H\sim 3\times10^{20}{\rm cm^{-2}}$, lower than $N_H>10^{22}{\rm cm^{-2}}$ observed in many IPs \citep{yu10}. With the distance $d=110$pc of AR~Sco, the column density   will be  mainly contributed  by the interstellar absorption. This also supports the hypothesis that
most of the X-ray emission of AR~Sco is not produced  as a result of the mass accretion on the WD surface, as we have discussed in section~\ref{timing}. 

Most of the  X-ray emission of the  AR~Sco is produced by the thermal plasma heated up to several keV. 
Since there is no evidence of the accretion column for AR~Sco, an alternative plausible process is  
a magnetically interaction between the WD and M-type star, such that  a magnetic dissipation process eventually heats up/accelerates 
the plasma on the M-star surface.  When the  WD's magnetic field lines sweep across the surface of the M-type star, 
 the magnetic  interaction on the M-type star produces an azimuthal component of WD's magnetic field, 
 and the pitch $\eta\equiv\delta B_{\phi}/B$ will increase at $\eta\rightarrow 1$ before the magnetic field becomes unstable against 
the dissipation  process.  We estimate the rate of the energy dissipation as  (\citealt{lai12,bu17})
\begin{eqnarray}
L_B&=&\frac{\eta B^2}{8\pi}(4\pi R_3\delta)\Omega_{WD}\sim 2.8\times 10^{32}{\rm erg~s^{-1}} \nonumber \\
&\times& \left(\frac{\mu_{WD}}{10^{35}{\rm G~cm^2}}\right)^2\eta\left(\frac{\delta}{0.01}\right)\left(\frac{R_*}{3\cdot 10^{10}{\rm cm}}\right)^3 \nonumber \\
&\times &\left(\frac{a}{8\cdot10^{10}{\rm cm}}\right)^{-6}\left(\frac{P_{s}}{117{\rm s}}\right)^{-1},
\end{eqnarray}
where $\mu_{WD}$ is the WD's magnetic dipole moment, $\delta$ is the skin depth \citep{bu17}, and  $\Omega_{s}=2\pi/P_{s}$ is the
spin angular frequency. Since the thermal component of the X-ray emission is observed with a luminosity of $\sim 4\times 10^{30}{\rm erg~s^{-1}}$, a small fraction of the dissipation energy is converted into keV plasma. Most of  the dissipation
energy would be used to accelerate the electrons, and be  radiated away by the synchrotron radiation that has a peak at
 the IR/optical bands in the spectral energy distribution and a synchrotron luminosity of 
$L_{syn}\sim L_{B}\sim 10^{32}{\rm erg~s^{-1}}$.

An evidence of the non-thermal emission can be found in the pulsed component, although the possibility of the emission 
from the thermal plasma cannot be ruled out.  However the alignment 
of the pulse peaks in  the  optical/UV/X-ray energy bands (three to four order of magnitude in the  energy)
strongly supports the  synchrotron emission process of  the non-thermal relativistic electrons.
The double peak structure in optica/UV bands  of AR~Sco is similar to the Crab pulsar (isolated young neutron star),  
for which the electrons/positrons are accelerated by the electric field parallel to the magnetic field line, where 
the charge density  deviates from the Goldreich-Julian charge density. As pointed out by \cite{ge16},  however, 
the number of the particles that emit the observed pulse optical emissions of AR~Sco is significantly larger than the
 number that can be supplied by the WD itself. This suggests that the synchrotron emitting electrons are supplied from the 
M-type star surface,  and the acceleration process is different from that of  NS pulsars.

Magnetic reconnection on the M-type star is a possible process  to
produce the relativistic electrons.  The strength of the magnetic field of the WD at the surface of the M-type star is  of the order of 
\begin{equation}
B_{d}\sim 195\left(\frac{\mu_{WD}}{10^{35}{\rm G~cm^3}}\right)\left(\frac{a}{8\cdot 10^{10}{\rm cm}}\right)^{-3}{\rm G}, 
\end{equation}
where $\mu_{WD}$ is the magnetic moment of the WD. In the observed SED (Figure~\ref{sed}), the spectral 
peak appears at $E_{p}\sim 0.1-1$eV.
If the observed pulse emission is originated from the M-type star surface, 
the synchrotron radiation implies
the typical Lorentz factor of 
$\gamma_e\sim 170(B_d/200{\rm G})^{-1/2}(E_{p}/1{\rm eV})^{1/2}$.
With the Lorentz factor $\gamma_e\sim 170$, on the other hand, the time scale of the synchrotron loss around
the M-type star is $\tau_s\sim 110{\rm s}(B_d/200{\rm G})^{-2}(\gamma_e/170)^{-1}$.
Since the cooling time scale is  longer than the crossing time scale of
$\tau_c\sim a/c\sim 2.5$s, the accelerated electrons on the M-type star surface can 
migrate into the  inner magnetosphere of the WD along the magnetic field of the WD before losing
their energy.

\cite{ta17} discuss  that  the observed pulse emission is produced by
the relativistic electrons trapped by the closed magnetic field lines of
the WD. The accelerated electrons from the M-type star will move toward
the WD's surface with the condition that $\tau_c<\tau_s$, and increase
the perpendicular momentum under the first adiabatic invariance.  The electron is rebounded
by the magnetic mirror effect and returns to outer magnetosphere. 
The synchrotron  emission from the first magnetic mirror point after leaving
the M-type star surface  dominates the emission
from the subsequent mirror points, and are observed as
the pulse emission (Figure~\ref{pulmodel}). In this scenario, the pulse emission
can be produced from  the inclined rotator, for which the dipole magnetic axis and the spin axis
are not aligned; in \cite{ta17}, 
the spin axis of WD is assumed to be  perpendicular to the orbital plane.
The electrons trapped into different magnetic field lines have different travel  time from
the M-type star surface to the first magnetic mirror point. Due to the difference in the travel times, the electrons injected at the different time  may
arrive the first magnetic mirror  point simultaneously. This enhances the observed flux and this effect becomes important for the electrons
that are injected around when the magnetic axis is laid within the plane made by the
spin axis and the direction of the M-type star.  Since the position of this plane relative to the direction
of the Earth shifts over  the orbital phase, the pulse peak also shifts in the spin phase, and results in
 the formation of the beat frequency in the timing analysis.

 In summary, we have reported that the X-ray emission from AR~Sco is modulating with
 the orbital phase and with the  beat phase. The X-ray orbital
 modulation  with a week absorption and synchronizing with the UV emission
 suggests that the most of the  emission is originated from
 the M-type star surface, rather than the WD's surface similar to other IPs.
 We found that  the pulse shape of the X-ray emission is similar to that in the optical/UV bands, and
 the peak position is aligned in the optical/UV/X-ray bands.
 This is strong evidence  that
 the pulse emission is the non-thermal, and  it is produced by the synchrotron
 radiation process of the relativistic electrons.  In the X-ray data,
 the  evidence of the non-thermal emission can be seen in the spectrum of
 the pulsed component. Our  results support  that AR~Sco is
 the new class of the  WD binary system that continuously produces
 the non-thermal radiation from the relativistic electrons.

 JT is supported by the National Science Foundation of China (NSFC)
 under 11573010, U1631103 and 11661161010.
 PHT is supported by NSFC through grants 11633007 and 11661161010.
 CYH is supported by the National Re search Foundation of Korea
 through grants 2014R1A1A2058590 and 2016R1A5A1013277.  AKHK is supported by
 the Ministry of Science and Technology of Taiwan through
 grants 105-2112-M-007-033-MY2, 105-2119-M-007-028-MY3, and 106-2918-I-007-005.  KSC  are supported by GRF grant under 17302315.

\end{document}